\documentstyle[preprint,epsf,aps]{revtex} 
\tightenlines
\begin{document}

\newcommand{\lsim}{\mathrel{\mathop{\kern 0pt \rlap
  {\raise.2ex\hbox{$<$}}}
  \lower.9ex\hbox{\kern-.190em $\sim$}}}
\newcommand{\gsim}{\mathrel{\mathop{\kern 0pt \rlap
  {\raise.2ex\hbox{$>$}}}
  \lower.9ex\hbox{\kern-.190em $\sim$}}}



\title{
On the Interpretation of the Atmospheric Neutrino Data in Terms
of Flavor Changing Neutrino Interactions}

\author{N.\ Fornengo, M.\ C.\ Gonzalez-Garcia, 
and J.\ W.\ F.\ Valle$^{1}$}
\address{\sl $^1$ Instituto de F\'{\i}sica Corpuscular -- C.S.I.C. \\
    Departamento de F\'{\i}sica Te\`orica, Universitat of Val\`encia \\
    46100 Burjassot, Val\`encia, Spain}
\maketitle
\vspace{.5cm}

\hfuzz=25pt

\begin{abstract} 
Flavour changing (FC) neutrino--matter interactions have been proposed
as a solution to the atmospheric neutrino anomaly. Here we perform the
analysis of the full set of the recent 52 kTy Super-Kamiokande
atmospheric neutrino data, including the zenith angle distribution of
the contained events as well as the higher energy upward--going
stopping and through-going muon events. Our results show that the FC
mechanism can describe the full data sample with a 
$\chi^2_{min}=44/33$ d.o.f which is acceptable at the 90 \% confidence 
level. The combined analysis 
confines the amount of FC to be either close to maximal or to the level of
about (10--50)\%.
\end{abstract}
\pacs{PACS numbers: xxx}

\newpage
 \section{Introduction}

Neutrinos produced as decay products in hadronic showers from cosmic
ray collisions with nuclei in the upper atmosphere ~\cite{flux} have
been observed by several detectors
\cite{Frejus,Nusex,Kamiokande,IMB,SuperKamiokande,Soudan}.  Although
the absolute fluxes of atmospheric neutrinos are largely uncertain,
the expected ratio $(\mu/e)$ of the muon neutrino flux ($\nu_\mu +
\bar{\nu}_\mu$) over the electron neutrino flux ($\nu_e+\bar{\nu}_e$)
is robust, since it largely cancels out the uncertainties associated
with the absolute flux.  In fact, this ratio has been calculated
~\cite{flux} with an uncertainty of less than 5\% over energies
varying from 0.1~GeV to 100~GeV. In this resides our confidence on the
long-standing atmospheric neutrino anomaly.

Although the first iron-calorimeter detectors in
Fr\'ejus~\cite{Frejus} and NUSEX~\cite{Nusex} reported a value of the
double ratio, R($\mu/ e$) = $(\mu/e)_{\rm data}/(\mu/e)_{\rm MC}$,
consistent with one, all the water Cerenkov detectors
Kamiokande~\cite{Kamiokande}, IMB~\cite{IMB} and
Super-Kamiokande~\cite{SuperKamiokande} have measured R($\mu/ e$)
significantly smaller than one.  Moreover, the Soudan-2 Collaboration,
also using an iron-calorimeter, reported a small value of R($\mu/
e$)~\cite{Soudan}, showing that the so-called atmospheric neutrino
anomaly was not a feature of water Cerenkov detectors.

Recent Super-Kamiokande high statistics
observations~\cite{SuperKamiokande} indicate that the deficit in the
total ratio R($\mu/ e$) is due to the number of neutrinos arriving in
the detector at large zenith angles.  Although $e$-like events do not
present any compelling evidence of a zenith-angle dependence, the
$\mu$-like event rates are substantially suppressed at large zenith
angles.

The $\nu_\mu \to \nu_\tau$ as well as the $\nu_\mu \to \nu_s$ 
\cite{atm98,atmo98} oscillation hypothesis provides a very good 
explanation for this smaller-than-expected ratio, which is also simple
and well-motivated theoretically. This led the Super-Kamiokande
Collaboration to conclude that their data provide good evidence for
neutrino oscillations and neutrino masses \cite{skos}.  However,
alternative explanations to the atmospheric neutrino data have been
proposed in the literature including the possibility of neutrino decay
\cite{decay}, the violation of relativity principles~\cite{vep,vli} 
or the violation of CPT symmetry \cite{vcpt}.  These explanations,
however, have been challenged by the precise data of Super-Kamiokande
on upward going muon events~\cite{upgoing} which allows to study the
energy dependence of the neutrino survival (or disappearance)
probability~\cite{lipari,lisivsr}. Based on such observations, both the
possibility of an explanation of the anomaly in terms of neutrino
decay \cite{lisidec} as well as the violation of relativity principles
or the violation of CPT symmetry \cite{lisivsr}, have been
disfavoured.

In Ref.~\cite{fcnc_1} an alternative explanation of the atmospheric
neutrino data in terms of FC neutrino-matter
interactions~\cite{fc_all} was proposed, and it was shown that even if
neutrinos have vanishing masses and/or the vacuum mixing angle is
negligible, FC neutrino matter interactions could account for the
Super-Kamiokande results on contained events providing an excellent
description to the data, statistically as good as neutrino
oscillations. The validity of this explanation was first questioned in
Ref. \cite{lipari} where the authors presented arguments against the
FC neutrino-matter interaction solution on the basis of a fit to the
up-going muons data from SuperKamiokande.

In this paper we re-analyze the possibility of explaining the
atmospheric neutrino anomaly by means of $\nu_\mu \to \nu_\tau$
conversion induced by flavour--changing neutrino--matter interaction
which can be effective during the neutrino propagation in the Earth.
We extend the analysis of Ref.~\cite{fcnc_1} to the new set of
Super-Kamiokande data by including also the up-going muon samples.

\section{Massless Neutrino Evolution with FC Interaction}

In our phenomenological approach we assume that the evolution
equations which describe the $\nu_\mu \to \nu_\tau$ transitions in
matter may be written as~
\begin{eqnarray} &  i{\displaystyle{d}\over 
\displaystyle{dr}}\left( 
\begin{array}{c} \nu_\mu 
\\ \nu_\tau  \end{array} \right) = \hskip .3cm & \hskip-.1cm
\sqrt{2}\,G_F \left( \begin{array}{cc} 0 &  \epsilon_\nu n_f(r)
\\ \epsilon_\nu n_f(r)& \epsilon_{\nu} ' n_f(r) \end{array} \right)
\left( \begin{array}{c} \nu_\mu  \\ \nu_\tau 
\end{array} \right) ,
\label{motion} 
\end{eqnarray}
where $\nu_a \equiv \nu_a (r)$ ($a=\mu,\tau$) are the probability
amplitudes to find these neutrinos at a distance $r$ from their
creation position, $\sqrt{2}\,G_F n_f(r) \epsilon_\nu$ is the
$\nu_\mu+ f \to \nu_\tau + f$ forward scattering amplitude and
$\sqrt{2}\,G_F n_f(r) \epsilon_\nu '$ is the difference between the
$\nu_\tau - f$ and $\nu_\mu - f$ elastic forward scattering
amplitudes, with $n_f(r)$ being the number density of the fermions
which induce such processes. 

The parameters $\epsilon$ and $\epsilon'$ contain the information
about FC neutrino interactions.  Such FC interactions may be
accompanied by neutrino mass~\cite{2227} but this need not be the
case~\cite{MV,FCSU5}. One description would be to parametrize directly
the FC interactions in terms of an effective four-fermion Hamiltonian.
This could, for instance, arise by renormalization effects from the
unification scale down to the electroweak scale in, say, supergravity
models~\cite{FCSU5}. An alternative more phenomenological way is to
consider the existence of a tree--level FC process $\nu_\alpha+ f \to
\nu_\beta + f$ where $f$ is an elementary fermion (charged lepton or
quark). The interaction can be mediated by a scalar or vector boson of
mass $m$ and the neutrino--fermion coupling is generically denoted by
$g_{\alpha f}$ ($\alpha$ is a flavour index) and can be written as
\begin{equation}
\epsilon^\prime_\nu = {|g_{\tau f}|^2 - |g_{\mu f}|^2 \over
4m^2\sqrt{2}\,G_F } \;\;\;\;\;\;
%
\mbox{and}\;\;\;\;\;\;
%
\epsilon_\nu   = {g_{\tau f} \cdot g_{\mu f} \over 4m^2\sqrt{2}\,G_F}. 
\label{epsilon}
\end{equation}

Since we are assuming vanishing neutrino masses, the anti--neutrino
transitions $\bar\nu_\mu\to \bar\nu_\tau$ are governed by the same
evolution matrix given in Eq.~(\ref{motion}).  For the sake of
simplicity, we consider $\epsilon_{\bar\nu}=\epsilon_\nu$ and
$\epsilon^\prime_{\bar\nu}=\epsilon^\prime_\nu$, which implies that we
have only two free parameters in the analysis. Moreover, we set our
normalization on these parameters by assuming that the relevant
neutrino interaction in the Earth is only with down-type quarks.  One
could also assume that the incoming atmospheric neutrino has FC
interactions off-electrons or equivalently, due to charge neutrality,
off-up-type quarks. For simplicity, in the present analysis we
consider only the case of interactions on down-type quarks.

We have calculated the transition probabilities of $\nu_\mu~ \to
\nu_\tau$ ($\bar \nu_\mu \to \bar\nu_\tau$) as a function of the
zenith angle by numerically solving the evolution equation using the
density distribution in \cite{PREM} and a realistic chemical
composition with proton/neutron ratio 0.497 in the mantle and 0.468 in
the core \cite{BK}.

\section{Fitting the Data to the FC Hypothesis}

We have then used these probabilities to compute, as a function of the
two parameters, $\epsilon_\nu$ and $\epsilon^\prime_\nu$, the
theoretically expected numbers of events for the four sets of data
reported by Super-Kamiokande: sub-GeV, multi-GeV, stopping muons and
through-going muons. The expected number of contained events are
computed by convoluting the probability with the corresponding
neutrino fluxes (for which we use the Bartol calculations
\cite{flux}) and interaction
cross sections and taking into account the experimental efficiencies
as detailed in Ref.~\cite{atm98}. For the up-going muon samples we
obtain the effective muon fluxes for both stopping and through-going
muons by convoluting the probabilities with the corresponding muon
fluxes produced by the neutrino interactions with the Earth. We
include the muon energy loss during propagation both in the rock and
in the detector according to ~\cite{muloss,ricardo} and we take into
account also the effective detector area for both types of events,
stopping and through-going.  We compute the effective area using the
simple geometrical picture given in Ref.~\cite{lipari1}. Our final
results show good agreement with the full MC simulation of the
Super-Kamiokande collaboration in the Standard Model case (see the
thick solid line in Fig. 3.)

In our statistical analysis we adopted the technique
\cite{atm98,fogli2} of fitting separately the angular distributions of
the $\mu$- and $e$-like contained events ($N^i_\mu$ and $N^i_e$, $i$
stands for sub-GeV and multi-GeV) and the up-going muon fluxes
($\Phi^j_\mu$, $j=$ stopping, through-going).  The expected number of
events have been compared with the recent 52 kTy data reported by the
Super-Kamiokande Collaboration \cite{taup} and the allowed regions in the
($\epsilon_\nu$, $\epsilon_\nu^\prime$) plane have been determined from a
$\chi^2$ fit.  In constructing the $\chi^2$ function, we explicitly
take into account the correlation of errors, both of theoretical and
experimental origin. Details on the definition of the correlation
matrix for contained events can be found in Ref.~\cite{atm98}, while
the definition of the sources of errors and their correlations for the
up-going muons fluxes are given in Ref.~ \cite{oscill,fogli2}.  Here
we simply summarize that we consider the overall normalization of the
up-going muon fluxes to be affected by an uncertainty of 20\% but in
order to account for the uncertainties in the primary cosmic ray flux
spectrum we allow a 5\% variation in the ratio between muon events in
different energy samples. We further introduce a 10\% theoretical
error in the ratio of electron-type to muon-type events of the
different samples.  Other important source of theoretical uncertainty
arises from the neutrino interaction cross section which at
Super-Kamiokande ranges from 10--15 \%.  Uncertainties in the ratio
between different angular bins are treated, similarly to
Ref.~\cite{fogli2}, by allowing a variation of 5\% times the
difference between the mean bin cosines.  With our definition we
obtain, for instance, $\chi^2_{SM}=$122/(35 d.o.f) which means that
the SM has a CL of $10^{-11}$ ! Using this same $\chi^2$ function for
the case of oscillations we obtained allowed regions for masses and
mixing angles very similar to those obtained by the Super-Kamiokande
collaboration both for contained events as well as for upward going
muons \cite{atm98}.

In Fig.~\ref{fig1} we show the contours of the regions allowed by
the Super-Kamiokande data. The different panels of the figure refer to
the fits performed over the different sets of data separately: (a)
sub-GeV; (b) multi-GeV; (c) stopping muons; (d) through-going
muons. The shaded areas are the regions allowed at 90\% C.L., while
the dashed and dotted contours refer to 95 and 99 \% C.L.,
respectively. The condition used to determine the allowed regions is:
$\chi^2 = \chi^2_{min} + \Delta \chi^2$ where $ \Delta \chi^2 = 4.6,
6.0, 9.2$ for 90, 95 and 99 \% C.\ L., respectively.

The allowed regions for the contained events are, as expected, 
similar to the ones obtained in Ref.~\cite{fcnc_1}. The individual
best fits now improve with respect to the analysis of the old data:
$\chi^2_{min} =$ 2.4/(8 d.o.f.) for the sub-GeV data ($\epsilon_\nu =
0.196$ and $\epsilon_\nu^\prime = 0.013$) and to $\chi^2_{min} =$
6.4/(8 d.o.f.) for the multi-GeV sample ($\epsilon_\nu = 0.689$ and
$\epsilon_\nu^\prime = 0.284$).  The combination of the two sets of
contained events leads to allowed regions which are analogous to the
ones reported in the Ref.~\cite{fcnc_1} and which are not reproduced
again here. The best fit point corresponds to $\epsilon_\nu =0.95$ and
$\epsilon_\nu^\prime = 0.084$ with $\chi^2_{min} =$ 9.3/(18 d.o.f.). 
The goodness of the fit to the contained events in the FC-neutrino interaction
scenario can be understood since the suppression of the expected event
rates for contained events is the same for sub-GeV and Multi-GeV
samples.  The Super--Kamiokande collaboration has also measured the
energy dependence of the up-down asymmetry for contained events
\cite{skos} and this clearly indicates a strong energy dependence of
the asymmetry for muon-like events in the momentum range 
$0.2$ GeV $<p_\mu< 2$ GeV. The asymmetry
is consistent with zero at low momentum but significantly deviates
from the expectation in the SM at higher momenta. One may naively
expect that since the FC conversion mechanism is energy-independent it
could be in contradiction with this measurement. However, one must notice
that the average angle between the directions of the final-state lepton
and the incoming neutrino ranges from $70^\circ$
at 200 MeV to $20^\circ$ at 1.5 GeV, so that at low momenta the
possible asymmetry of the neutrino flux is largely washed out.  
In Fig.~\ref{fig2} we plot, together with the
Super--Kamiokande data, the momentum behaviour of the asymmetry
in the FC-neutrino interaction scenario calculated for the best fit 
point to the contained event sample. 
As seen in the figure the agreement is excellent.

In Fig.~\ref{fig1}, we also show the regions which are allowed
by the up-going muons samples of Super-Kamiokande. Panel (c) stands for
stopping muons and panel (d) for the through-going sample.
In the case of stopping muons, we see that, analogously to the
contained events, the allowed region lies in the sector of the plane
where the average survival probability is of the order of a half,
which is what appears to be needed for explaining the data. Instead,
in the case of through-going muons, the experimental data do not show
such a strong reduction with respect to the theoretical calculations,
and therefore the allowed region lies in the upper-left corner of
the parameter space, which refers to a smaller transition probability.
In both cases, the best fit point for each individual
sample is good: $\chi^2_{min} =$ 1/(3 d.o.f.) for stopping
muons ($\epsilon_\nu = 0.756$ and $\epsilon_\nu^\prime = 0.196$) and
$\chi^2_{min} =$ 10.3/(8 d.o.f.) for the through-going case
($\epsilon_\nu = 0.081 $ and $\epsilon_\nu^\prime = 0.260$).  Both for
the contained and for the up-going events, the best fits have the same
level of statistical confidence as compared to the oscillation
interpretation of the atmospheric neutrino data. This is shown in
Table I, where we report the best fit values we obtain for the
different data sets in the case of the FC--$\nu$ interactions scenario
and in the case of the neutrino oscillation scenario\cite{oscill}.

The allowed regions
can be qualitatively understood in the approximation of constant
matter density. The conversion probability in this case is
\begin{equation}
P(\nu_\mu \to \nu_\tau)= 
\frac{4\epsilon_\nu^2}{4\epsilon_\nu^2+{\epsilon_\nu^\prime}^2}
\sin^2({1\over 2} \eta L), 
\label{Eq:prob}
\end{equation}
\noindent
where $\eta = 
\sqrt{4\epsilon_\nu^2+{\epsilon^\prime_\nu}^2} \sqrt{2} G_F n_f$. 
For $n_f = n_d\approx 3n_e$ and 
$\epsilon^\prime_\nu < \epsilon_\nu$, the oscillation 
length in matter is given by 
\begin{equation}
L_{osc} = \frac{2\pi}{\eta} \approx 1.2\times 10^3 
\left[ \frac{2\ \mbox{mol/cc}}{n_e} \right]
\left[ \frac{1}{\epsilon_\nu} \right] \ \mbox{km}.
\label{osclength}
\end{equation}

{}From Eq.~(\ref{Eq:prob}) one can see that in order to have a 
relatively large transition probability, as required by the
contained events and, also, by the stopping muons events,
the FC parameters are required to be in the region $\epsilon^{\prime}_\nu
\lesssim \epsilon_\nu$ and $\eta\gtrsim \pi/R_{\oplus}$. This last
condition leads to a lower bound on $\epsilon_\nu$.  The island in
Fig.~\ref{fig1}.(b) corresponds to $\eta\sim\pi/R_{\oplus}$.

The combination of the different data sets in a single
$\chi^2$-analysis is shown in Fig.~\ref{fig3}. Panel (a) shows the combination
of the full angular distribution of contained events with the total
(unbinned) event rate of stop and through-going muons data, while
panel (b) refers to the combination of all the angular distributions,
including that of through-going muon events. In Fig.~\ref{fig3}.(a) the
information brought by the higher energy data is effective at the
normalization level, since no information about their angular
dependence is included. In this case the allowed region is still relatively
large although the description is already worse than in the oscillation
case as can be seen by comparing the corresponding $\chi^2_{min}$
(20.4/(10 d.o.f.) for FC as compared to 9.6/(10 d.o.f.) for the oscillation scenario).
This worsening is due to the fact that in the FC
scenario the transition probability is energy independent while the
data shows a smaller conversion for the higher energy through-going
muon events. As seen in Fig.~\ref{fig3}.(b), when the angular information 
of both
stopping and through-going muons is included in the data analysis, the
description becomes even worse, mainly due to the angular 
distribution of the through-going data set. The allowed regions now 
form a set of isolated small 'islands'. The best fit point corresponds to 
$\epsilon_\nu = 0.57$ and $\epsilon_\nu^\prime = 0.45$
acceptable at the 90 \% CL ($\chi^2_{min} =$ 44/(33 d.o.f.)). 

The behavior of the allowed regions can be understood by observing
Fig.~\ref{fig4} where we show the angular distributions for the four cases: (a)
sub-GeV; (b) multi-GeV; (c) stopping muons; (d) through-going
muons. We show the distributions for the best fit point obtained from
the combination of contained events with the total number of upward
going  muons $P_1 = (\epsilon_\nu,\epsilon_\nu^\prime) $
= (0.17,0.28) and for the best fit point obtained from the analysis
of the full data set $P_2$ = (0.57,0.45).  Although both points give a
similar normalization to the up-going muon data samples, point $P_1$
gives a better description to the angular dependence of the contained
events, but it does not describe well the zenith angle distribution of
the through-going muon events. As commented above, such point
correspond to an effective FC-oscillation length of the order of the
Earth radius. In this case we can see the imprints of the
``oscillatory'' sine behavior in the expected angular distribution of
the up-going muon events. Such behavior, however, does not
appear to be present in the Super-Kamiokande data, leading to a worse
overall fit.  In the case of multi-GeV contained events, this
oscillatory behavior is averaged out due to the smaller angular
resolution in the data and point $P_1$ can give a good description
of the data. On the other hand, point $P_2$ gives a worse description
of the contained events but fits better the shape of upward going muon data,
with the exception of the last three angular bins of the
through--going sample, where it does not
produce a sufficient amount of through-going muons
at angles $0<\theta < 20^\circ$ below the horizon.

\section{Discussion}

What can we say about the required strength of the neutrino-matter
interaction in order to obtain a good fit of the observed data?
{}From our best fit results we obtain that the atmoshperic
neutrino data can be explained if $\epsilon_\nu \gsim 0.4$ 
and $\epsilon'_\nu \gsim 0.1$. In terms of the neutrino--quark
couplings introduced in Eq.~(\ref{epsilon}) we see that for
masses $m\approx 200$ GeV combinations of couplings
$g_{\tau f} \cdot g_{\mu f}$ and $|g_{\tau f}|^2 - |g_{\mu f}|^2$ 
of the order of 1 are needed.

Model independent constraints on the $\epsilon$
parameters can be extracted from their contribution to the $\nu_\mu$
neutral current cross section measured at low energy \cite{CCFR}. These
limits are stronger for interactions with quarks due to the better
precision of the $\sigma^{\nu_\mu}_{NC,N}$ data as compared to the
$\sigma^{\nu_\mu}_{NC,e}$.  We have estimated that these data constrain
$\epsilon \lesssim {\cal O}(0.1)$--${\cal O}(1)$, depending on the
fermion $f$ coupled to the neutrino. No limit on the $\epsilon^\prime$
parameter can be obtained from these measurements.
From this, we see that the FC mechanism is somewhat disfavoured,
although not strictly ruled out.
Additional limits can be obtained from the non-observation of 
lepton flavour violation in $\tau$ decays. Indeed, if $SU_L(2)$ 
symmetry is assumed~\cite{taudec} these limits are rather stringent.
However, they strongly depend on
the amount of $SU_L(2)$ violation present in the model.  For the
purpose of illustrating this explicitly, let us consider 
a supersymmetric model with broken $R$-parity as a way to
parameterize the FC neutrino-matter interaction~\cite{Ross:1985yg}. In
this case the FC $\nu_\mu$-matter interactions are mediated by a
scalar down-type quarks, $\tilde{d_j}$, so that we need only to check
the couplings where a $d$-quark and a $\mu$- or $\tau$-neutrino is
involved,{\it i.e} $g_{id} \approx \lambda^\prime_{ij1}$, $i=2,3$.
The $\lambda_{ijk}'$ are the coupling constants in the broken
$R$-parity superpotential $\lambda_{ijk}' L_iQ_jD^c_k$, where $L$, $Q$
and $D$ are standard superfields, and $4 \sqrt{2} G_F \epsilon_\nu =
|{\displaystyle \sum_j}\lambda_{3j1} '\lambda_{2j1}
'/\tilde{m}_{\tilde d j}^2|$.  The most stringent limit to the values
of the relevant FC quantities comes from limits on the FC tau decay
BR($\tau^-\to \rho^0+\mu^-) < 6.3 \times 10^{-6}$~ which implies that
$|{\displaystyle \sum_j}\lambda_{3j1}' \lambda_{2j1} ' (100$
GeV$/\tilde{m}_{\tilde u j})^2 | \lsim 3.1 \times 10^{-3}$.  This
constraint can be satisfied by cancellation between the contributions
from the up-squarks exchange of the third and second generation, although a
certain degree of fine-tuning is needed between their masses and
couplings. Notice that the $\epsilon_\nu$ may still be large as long as
the cancellation does not occur for the down-quarks.  This  can be
achieved by a small splitting between the up and down-squark masses
without conflicting with the limits from $\Delta\rho$.

\section{Conclusions}

In summary, in this paper we have re-analyzed the possibility of explaining the
atmospheric neutrino anomaly by means of $\nu_\mu \to \nu_\tau$
conversion induced by flavour--changing neutrino--matter interaction
which can be effective during the neutrino propagation in the Earth.
We extend the analysis of Ref.~\cite{fcnc_1} to the new set of
Super-Kamiokande data by including also the up-going muon samples.
Our results show that that flavour changing $\nu_{\mu}$-matter
interactions are able to describe the full set of data of 
Super-Kamiokande on atmospheric neutrinos at the 90 \% CL.
The agreement between the data and the calculated events for
the Super-Kamiokande detector is notably good for the
individual sets of data collected by Super-Kamiokande,
with a confidence level as good as for the oscillation hypothesis.
When the data are combined together, in particular once the 
upward--going muon zenith--angle distribution is included in the analysis, 
the $\nu_\mu$
oscillation provides a much better description. However, the
FC mechanism is compatible with the data at the quoted 90\% level
of statistical confidence.
The worsening of the fit which occurs when the through--going
muons sample is included is partly due to the fact that in the FC
scenario the transition probability is energy independent while the
data shows a smaller conversion for the higher energy up-through-going
muon events. The ensuing result is that the expectations
from FC-neutrino interaction for neutrinos arriving mainly at angles 
above 20 degrees below the horizon do not reproduce the experimental data.

The amount of FC neutrino interactions required
by our combined analysis in order to fit the data
is somewhat large, either close to maximal or at 
the level of about (10-50)\%. Although significant,
FC at this level in the neutrino sector can be accomodated 
in speficic models without conflicting with existing
experimental limits. These realizations of large
neutrino FC may require some degree of fine tuning,
like for instance in broken R-parity supersymmetric models,
but are theoretically viable.

The above FC mechanism can be also tested at future Long Baseline
experiments. {}From Eq.~(\ref{Eq:prob}), using $n_e \sim 2$ mol/cc, we
can predict that for $\epsilon \simeq \epsilon^\prime \sim 1 $ (0.1) the
planned K2K experiment ~\cite{K2K} should obtain $P(\nu_\mu \to
\nu_\tau) \sim 0.35$ (0.004) while for MINOS~\cite{MINOS} one finds 
$P(\nu_\mu \to \nu_\tau) \sim 0.75$ (0.04).

In conclusion, we have to comment that although FC
interactions would be eventually ruled out as the {\sl only} source of the
modification of the atmospheric neutrino predictions with respect to
those of the Standard Model they could still be there at
some level, even if
the data would admit a very good interpretation in terms of standard
$\nu_\mu \rightarrow \nu_\tau$ oscillations.
This is theoretically not an {\em ad hoc} assumption, since
in many theoretical models neutrino masses naturally
co-exist with FC-neutrino interactions. 

\begin{figure}[t]
\centering\leavevmode
\epsfxsize=\hsize
\epsfbox[10 40 530 555]{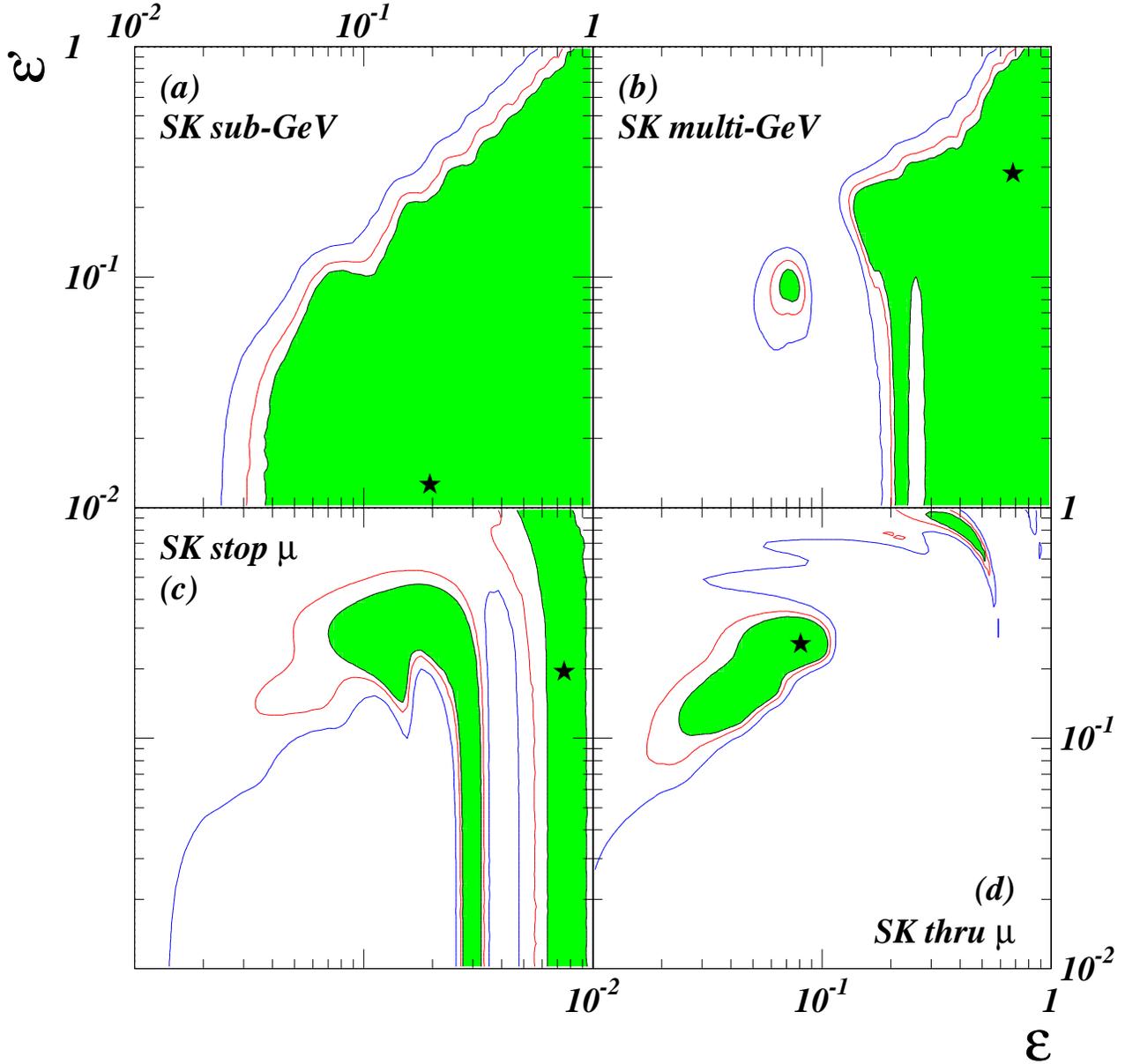}
\caption{Allowed regions for $\epsilon_\nu$ and $\epsilon_\nu^\prime$ 
in the FC massless-neutrino scenario for the different
Super-Kamiokande data sets: (a) sub-GeV, (b) multi-GeV, (c) stopping muons
and (d) through-going muons. 
The best fit points for each case are indicated by stars. The shaded
area refers to the 90\% C.L.. while the contours stand for 95\% 
and 99\% C.L.}
\label{fig1}
\end{figure}

\begin{figure}[t]
\centering\leavevmode
\epsfxsize=\hsize
\epsfbox[77 270 470 540]{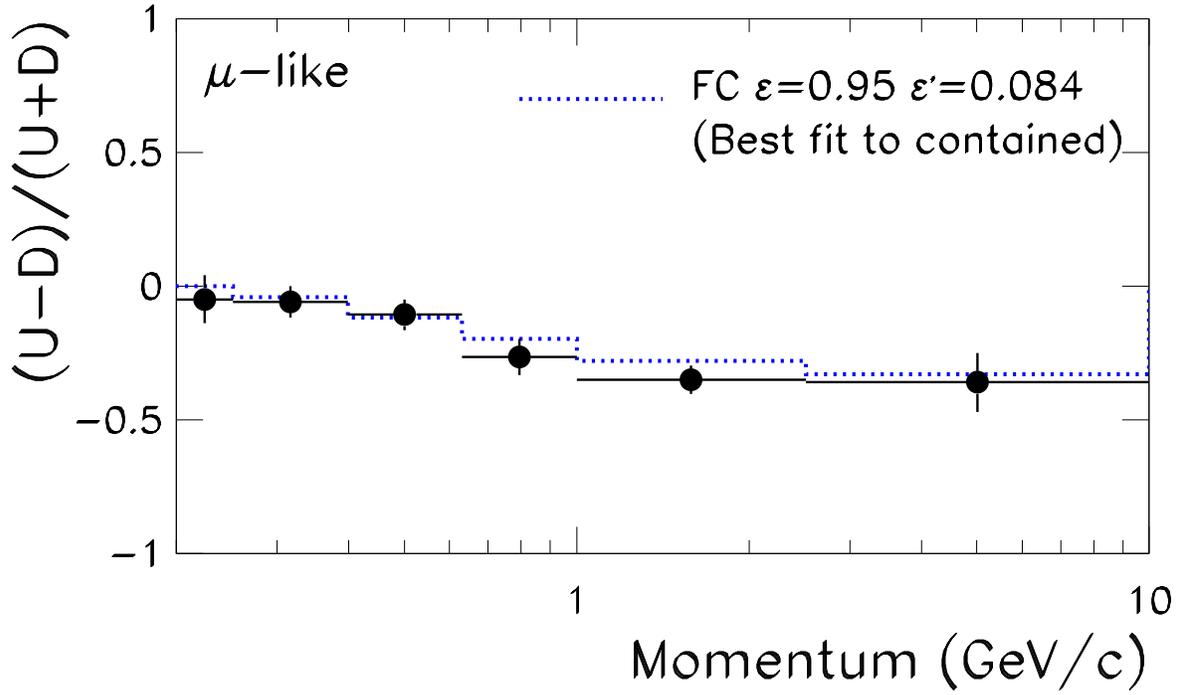}
\caption{ Expected up-down asymmetry in Super--Kamiokande
for the FC-neutrino scenario as a function of the muon momentum 
for fully contained $\mu$-like events, compared to the
Super--Kamiokande experimental data.}
\label{fig2}
\end{figure}

\begin{figure}[t]
\centering\leavevmode
\epsfxsize=\hsize
\epsfbox[20 250 530 555]{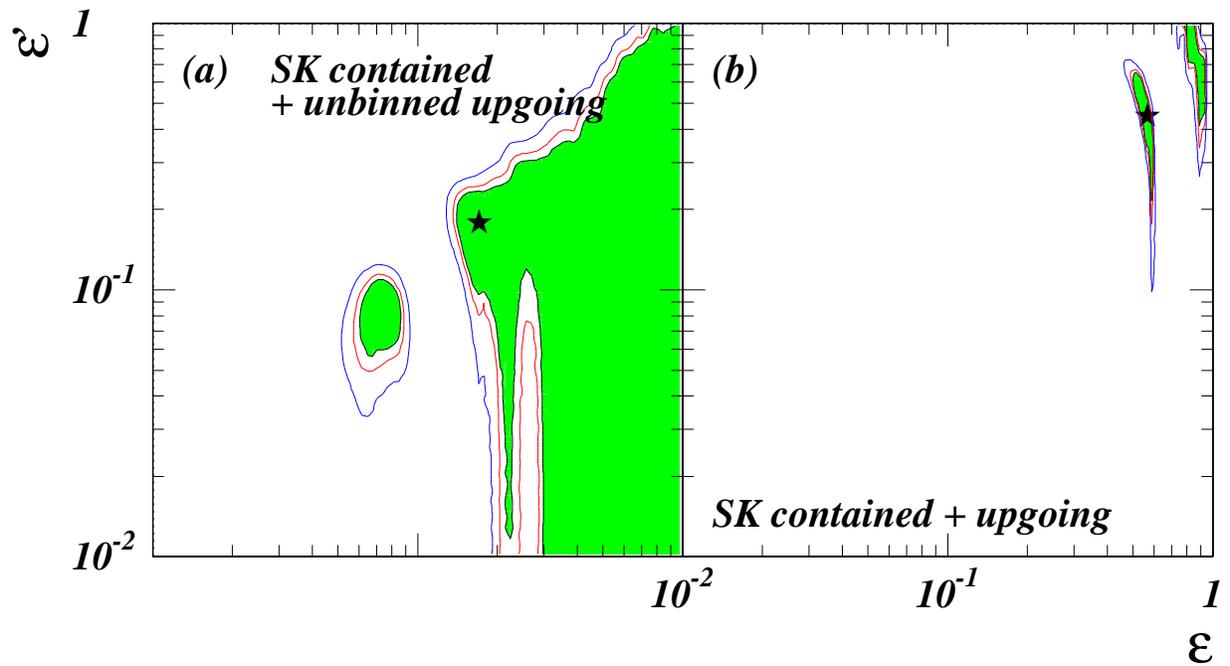}
\caption{Allowed regions for $\epsilon_\nu$ and $\epsilon_\nu^\prime$
for the combination of the Super-Kamiokande data sets: (a) the binned
contained events are combined with total (unbinned) up-going events;
(b) binned contained and up-going events. The best fit points for each
case are indicated by stars. The shaded area refers to the 90\% C.L.
while the contours stand for 95\% and 99\% C.L.}
\label{fig3}
\end{figure}

\begin{figure}[t]
\centering\leavevmode
\epsfxsize=12cm
\epsfbox[25 35 520 775]{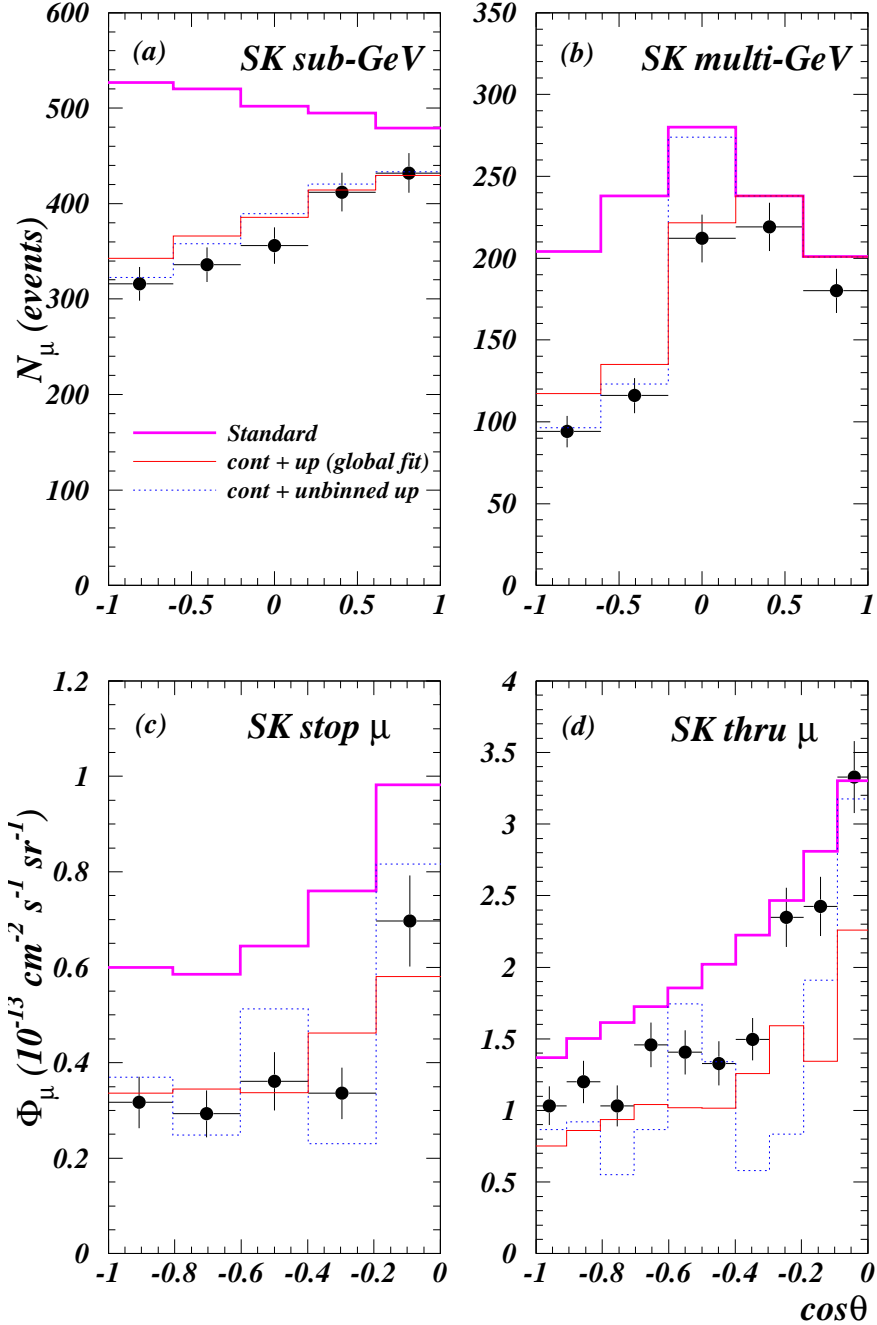}
\caption{Best-fit zenith angle distributions in the massless-neutrino
FC scenario. The thick-solid lines correspond to the calculation in
absence of new physics. The dotted lines correspond to the best
fit point obtained by the analysis of the contained events combined 
with total (unbinned) up-going events. The thin-solid line is for the
best fit point of the combined analysis of contained and up-going
muon events. The 52 kTy Super-Kamiokande data are
indicated by crosses.}
\label{fig4}
\end{figure}

On the other hand, one may turn the argument the other way around: should 
the atmospheric neutrino anomaly be explained in terms of neutrino
oscillations, then it will be possible to use the non-observation of
an additional effect in the atmospheric neutrino data in order to
impose new model independent limits on the strength of the $\epsilon$
and $\epsilon^\prime$ parameters, as can be foreseen by looking at
Fig.~\ref{fig1}(d).


\acknowledgements 

We thank R. Vazquez for providing us with the program to compute the
muon energy loss. We are also grateful to S. Nussinov for pointing
us out the limits from the neutrino neutral current measurements.
M.C. G-G is grateful to the Instituto de Fisica
Teorica from UNESP and to the CERN Theory division for their kind
hospitality during her visits.  This work was supported by Spanish
DGICYT under grant PB95-1077 and by the European Union TMR network
ERBFMRXCT960090.


\newpage
\begin{table}
\begin{tabular}{|l|r|r|r|}
Data & d.o.f & $\chi^2_{min FC}$ & $\chi^2_{min Osc}$ \\ 
\hline
sub-GeV            &  8  &  2.4  &  2.4  \\
multi-GeV          &  8  &  6.4  &  6.3 \\
contained          & 18  &   9.3  &  8.8  \\
stopping-$\mu$     &  3  &   1.   &  1.3     \\
through-going-$\mu$    &  8  &   10.3 &  10.4  \\
contained + total up-$\mu$     & 10  & 20.4  & 9.6 \\
contained + angular  up-$\mu$  & 33  & 44.  &  23.5 \\
\end{tabular}
\caption{$\chi^2_{min}$ obtained for several data combinations 
in the framework of FC--$\nu$ interactions  as compared to the case 
of the neutrino vacuum-oscillation scenario.}
\end{table}

\newpage
\newpage

\newpage

\newpage

\end{document}